\documentclass[prl,twocolumn,aps,superscriptaddress]{revtex4}
\usepackage{graphicx}

\begin{document}

\title{Explosive first order transition to synchrony in networked chaotic oscillators}

\author{I. Leyva}
\affiliation{Complex Systems Group, Univ. Rey Juan Carlos, 28933 M\'ostoles, Madrid, Spain}
\affiliation{Center for Biomedical Technology, Univ. Polit\'ecnica de Madrid, 28223 Pozuelo de Alarc\'on, Madrid, Spain}
\author{R. Sevilla-Escoboza}
\affiliation{Centro Universitario de Los Lagos, Univ. de Guadalajara, Lagos de Moreno, Jalisco 47460, Mexico}
\author{J. M. Buld\'u}
\affiliation{Complex Systems Group, Univ. Rey Juan Carlos, 28933 M\'ostoles, Madrid, Spain}
\affiliation{Center for Biomedical Technology, Univ. Polit\'ecnica de Madrid, 28223 Pozuelo de Alarc\'on, Madrid, Spain}
\author{I. Sendi\~na-Nadal}
\affiliation{Complex Systems Group, Univ.  Rey Juan Carlos, 28933 M\'ostoles, Madrid, Spain}
\affiliation{Center for Biomedical Technology, Univ. Polit\'ecnica de Madrid, 28223 Pozuelo de Alarc\'on, Madrid, Spain}
\author{J. G\'omez-Garde\~nes}
\affiliation{Dep. de F\'{\i}sica de la Materia Condensada, Univ. de Zaragoza, Zaragoza 50009, Spain}
\affiliation{Institute for Biocomputation and Physics of Complex Systems (BIFI), Univ.  de Zaragoza, 50009 Zaragoza, Spain}
\author{A. Arenas}
\affiliation{Institute for Biocomputation and Physics of Complex Systems (BIFI), Univ.  de Zaragoza, 50009 Zaragoza, Spain}
\affiliation{Dep. d'Enginyeria Inform\`{a}tica i Matem\`{a}tiques, Univ. Rovira i Virgili, 43007 Tarragona, Spain}
\author{Y. Moreno}
\affiliation{Institute for Biocomputation and Physics of Complex Systems (BIFI), Univ. de Zaragoza, 50009 Zaragoza, Spain}
\affiliation{Dep. de F\'{\i}sica Te\'orica, Univ. de Zaragoza, Zaragoza 50009, Spain}
\author{S. G{\'o}mez}
\affiliation{Dep.  d'Enginyeria Inform\`{a}tica i Matem\`{a}tiques, Univ.  Rovira i Virgili, 43007 Tarragona, Spain}
\author{R. Jaimes-Re\'ategui}
\affiliation{Centro Universitario de Los Lagos, Univ. de Guadalajara, Lagos de Moreno, Jalisco 47460, Mexico}
\author{S. Boccaletti}
\affiliation{Center for Biomedical Technology, Univ.  Polit\'ecnica de Madrid, 28223 Pozuelo de Alarc\'on, Madrid, Spain}

\begin{abstract}
Critical phenomena in complex networks, and the emergence of dynamical abrupt transitions in the macroscopic state of the system are currently a subject of the outmost interest. We report evidence of an explosive phase synchronization in networks of chaotic units. Namely, by means of both extensive simulations of networks made up of chaotic units, and validation with an experiment of electronic circuits in a star configuration, we demonstrate the existence of a first order transition towards synchronization of the phases of the networked units. Our findings constitute the first prove of this kind of synchronization in practice, thus opening the path to its use in real-world applications.


\end{abstract}

\maketitle

The understanding of the spontaneous emergence of collective behavior in ensembles of networked dynamical units constitutes a fascinating challenge in science. Despite the fact that critical phenomena in networks have been intensively studied, the physics literature \cite{doro08} almost exclusively reports continuous phase transitions. However, it has been recognized that, although very few in number \cite{echenique05,karsai07,prl}, there are physical processes which might lead to sharp, discontinuous transitions of a global order parameter. The last several years have also witnessed an ever-increasing interest in studying networked systems composed of nonlinear dynamical units \cite{physrepbocca}, and in particular, in the emergence of synchronization phenomena \cite{physreparenas}. Within this latter context, some advances have been made for the case of non-equilibrium synchronization transitions of chaotic systems \cite{tessone06,cencini08}, being, however, all the reported cases examples of second order phase transitions.

More recently it has been shown that discontinuous transitions can take place in networks of periodic oscillators \cite{prl}. There, a first-order non-equilibrium synchronization transition has been proved in scale-free networks, as a consequence of a positive correlation between the heterogeneity of the connections and the natural frequencies of the oscillators. Whether such an explosive behavior is restricted to periodic oscillators, or can be generalized to more complicated dynamical units remained an open problem.

In this Letter, we investigate the critical properties of synchronization transitions in heterogeneous networks of chaotic units (when the aforementioned interplay between topology and dynamics is taken into account), and we will give the first numerical and experimental evidence of an explosive phase synchronization in such ensembles of complex systems.
Even though the extent of our discussion is valid regardless of the specific phase coherent \cite{physrepsincro} dynamics that is considered for the evolution of each network's node, from now on we will focus on a specific benchmark chaotic system. Additionally, our choice is dictated by the need of implementing a qualitatively similar model in the laboratory, and therefore by the unique opportunity of contrasting the numerical predictions with the experimental evidence.

Let us consider an ensemble of $N=1,000$ piecewise R\"ossler units, interacting in a network via a bidirectional diffusive-like coupling \cite{Pisarchik06,Pisarchik09}:
\begin{eqnarray}
\dot{x_i} &=& -\alpha_i \left[ \Gamma \left( x_i-d\sum_{j=1}^N a_{ij}(x_j-x_i) \right) + \beta y_i + \lambda z_i \right] \nonumber \,, \\
\dot{y_i} &=& -\alpha_i(- x_i + \nu y_i) \,, \\
\dot{z_i} &=& -\alpha_i(-g(x_i)+z_i) \nonumber \,,
\label{modelo}
\end{eqnarray}
where the piecewise part is:
\begin{equation}
g(x_i)=\left\lbrace \begin{array}{cc}
0 & \mbox{if $x_i\leq 3$} \\
\mu(x_i-3) & \mbox{if $x_i > 3$}
\end{array} \right. \ .
\end{equation}

Every node (indexed by $i=1,\ldots,N$) is here represented by an associated three-dimensional vector ${\bf x}_i (t) \equiv (x_i(t), y_i(t), z_i(t))$. The parameters are: $\Gamma=0.05$, $\beta =0.5$, $\lambda=1$, $\mu = 15$,  and $\nu = 0.02-\frac{10}{R}$, where $R$ is a tunable quantity that regulates the dynamical state of the system. In particular, $R$ induces a chaotic dynamics\cite{Pisarchik06} in the range $R = [55, 110]$. ${\bf A} = \{a_{ij}\}$ is the adjacency matrix ($a_{ij}=1$ if units $i$ and $j$ are connected, and 0 otherwise), and $d$ is the coupling strength. Notice that the coupling is here applied through the $x$ variable, but equivalent results are obtained using a coupling in the variable $y$.

Finally, the natural oscillation frequency of node $i$ depends linearly on the parameter $\alpha_i$. Therefore, as long as one is concerned with imposing a positive correlation between the natural frequency and the degree $k_i$ of each unit (the number of connections that the $i^{th}$ unit is forming with the rest of the network), the $\alpha_i$ values (and therefore, the oscillators' frequencies) are distributed following the relation:
\begin{equation}
\alpha_i=\alpha \left(1+\Delta \alpha \frac{k_i-1}{N}\right)\,,
\label{corre}
\end{equation}
where $\alpha=10^4$, and $\Delta \alpha$ is a factor that determines the slope of the linear distribution. As a consequence, all nodes with degree $k=1$ have the same natural frequency regardless of the value of $\Delta \alpha$.  On the other hand, the range of frequencies in the ensemble becomes wider as $\Delta \alpha$ and $k_{\max}$ (the maximum degree in the network) are increased, i.e.\ as more heterogeneous degree distributions and/or steeper slopes are considered.
\begin{figure}
\includegraphics[scale=0.65]{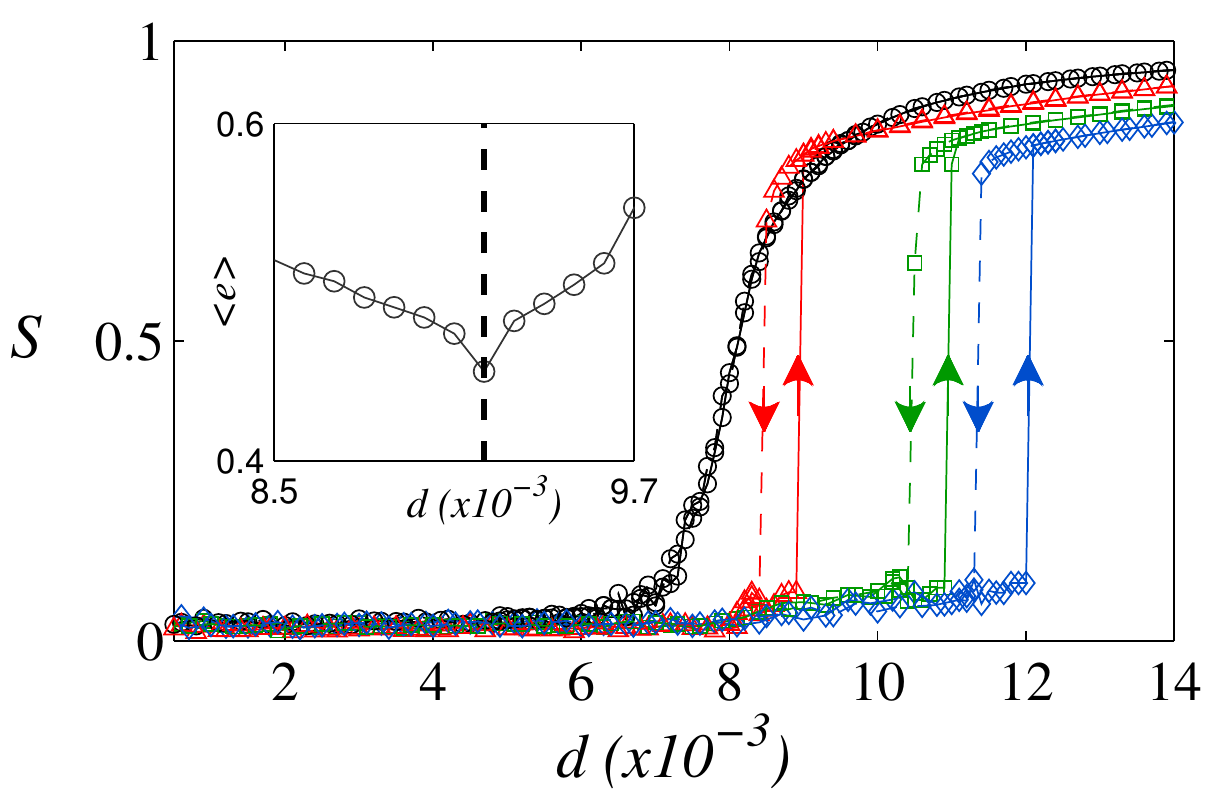}
\caption{(Color online). Phase  synchronization degree $S$ as a function of the coupling strength $d$ for different SF networks of size
$N=1,000$, and average degree $\langle k\rangle = 6$. The networks are built using the
configuration model \cite{conf} (see text for details). $\gamma=2.2$ (red
triangles), $\gamma=2.5$ (green squares), $\gamma=3.0$ (blue diamonds
and black circles). The correlation between node degree and natural
frequency is set via Eq. (\ref{corre}). $\Delta\alpha=6.0$ and $R=70$ for
the black circles case, while $\Delta\alpha=10.0$ and $R=100$ for the other
networks. Continuous (dashed) lines mark the forward (backward)
simulations, as $d$ is increased (decreased) in steps of $\delta
d=3 \times 10^{-4}$. The inset plot reports the average synchronization error $\langle e \rangle$ vs.~$d$ in the
proximity of the first-order transition occurring for $\gamma=3.0$, $R=100$ and
$\Delta\alpha=8.0$.}
\label{figurina1}
\end{figure}

The state of the network is monitored as a function of the coupling, by gradually increasing the value of $d$ in steps $\delta d$ along the simulation from $d=0$ (without resetting the system, as it will be done later in the experiment). Whenever the coupling is increased in $\delta d$, a long transient is discarded before the data are used for further processing. Furthermore, as long as we are looking for a first-order phase transition (and thus for an expected associated synchronization hysteresis), we perform the simulations also in the reverse way, i.e.\ starting from a given value $d_{\max}$ where the ensemble is phase synchronized, and gradually decreasing the coupling by $\delta d$ at each step. In what follows, the two sets of numerical trials are termed as {\em forward} and {\em backward}, respectively.

We here focus in the phase synchronization regime \cite{physrepsincro,kurths}. For each node $i$, the instantaneous phase at time $t$ is geometrically evaluated \cite{0sipov97} as $\phi_i(t) = \arctan \left( y_i(t)/x_i(t) \right)$, and the mean synchronization degree is calculated as $S = \left\langle \left| \frac{1}{N}\sum_{j=1}^{N} e^{i\phi_j(t)} \right| \right\rangle_t$, where the vertical bars denote the module and the angle brackets a temporal averaging.

The first result is that the delicate equilibrium between the network heterogeneity and the frequency distribution plays a crucial role in determining the nature of the phase synchronization transition. This is shown in Fig.~1, where we compare the dependence of the synchronization degree $S$ on the coupling strength $d$, in different networks of same size and average degree. All graphs are scale-free (SF) networks, obtained with the configuration model \cite{conf}, and featuring a degree distribution $P(k)\sim k^{-\gamma}$, with $\gamma=2.2$ (red triangles), $\gamma=2.5$ (green squares), and $\gamma=3.0$ (blue diamonds and black circles). Remarkably, the specific chaotic state of the nodes is a relevant feature, as demonstrated by comparing the case $\Delta\alpha=6.0, R=70$ (black circles, second order phase transition) with the case
$\Delta\alpha=10.0, R=100$ (blue diamonds, first order phase transition) for $\gamma=3.0$. The explosive (first order) character of the transition is clearly manifested by the presence of a hysteresis region, delimited in the figure by the continuous and dashed lines that mark, respectively, the forward and backward cycles. The inset plot shows the average synchronization error $\langle e \rangle$ (the average of the Euclidean distance between all pairs of nodes, normalized to account for the dependence of the mean amplitude oscillations on the parameters) vs.~$d$, in the proximity of the transition for $\gamma=3.0$, $R=100$ and $\Delta\alpha=8.0$.

The conclusive information conveyed by Fig.~1 is that the collective behavior emerging in the system is, indeed, a synchronization of the {\it phases} of the chaotic units, which is not associated to any specific amplitude correlation (as the error $\langle e \rangle$ is even increasing at the right side of the transition point).

A more exhaustive description of the phenomenon emerges from the exploration of the nature of the transition in the full parameter space $d-R$. The results are shown in Fig.~2, where we report (in color code) the values of $S$, for both forward and backward simulations. As it can be seen, the plane $d-R$ can be clearly divided in two areas (denoted as $I$ and $II$ in the figure) where the transition is of the first and second order, respectively. The striped portion of the area where the transition is of the first order marks the region where the hysteresis phenomenon is observed.

 A first conclusion of this Letter is that our numerical study reveals the presence, as a generic feature, of a first-order non-equilibrium phase transition towards phase synchronization in a network of chaotic oscillators. The phenomenon is here originated by a setting of the constituent nodes within a proper chaotic regime, together with imposing a positive correlation between the heterogeneity of the graph connections and the natural frequencies of the oscillators. It is worth noticing that in principle one can have first-order transitions for more homogeneous topologies, however at the cost of expanding the range of natural frequencies in the network.

\begin{figure}
\includegraphics[width=60mm,height=60mm]{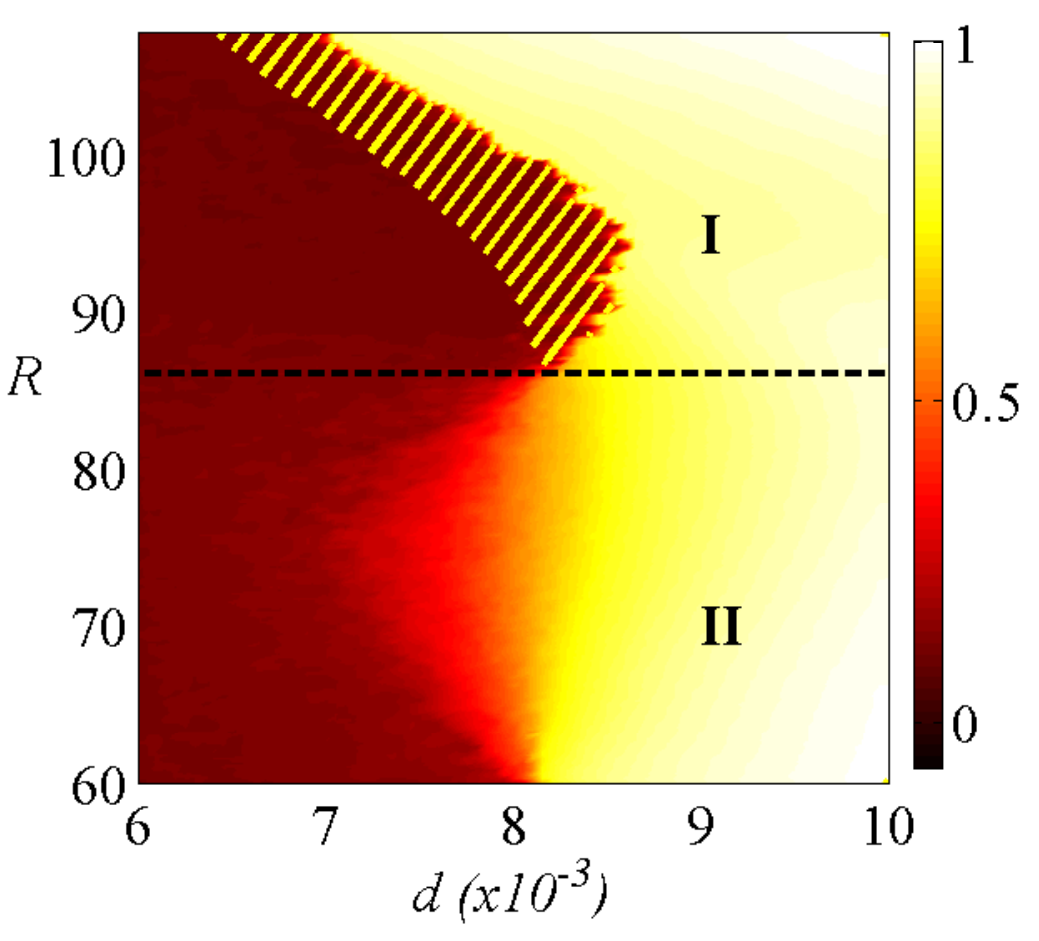}
\caption{(Color online). Mean synchronization degree $S$ (see text for definition)
in the parameter space $d-R$ (the values of $S$ are color-encoded, according 
to color-bar at the right of the panel).
The horizontal dashed
line marks the separation between the region of the parameter space where a second-order transition occurs
(below the line), and that for which the transition is instead of the first-order type (above the line).
In this latter region, a typical hysteresis phenomenon in the forward and
backward simulations is observed (whose extent is here marked by a striped area).
The step used for the variations in the coupling strength is 
$\delta d=5 \times 10^{-5}$. The entire phase diagram refers to the case of a
scale free network with $N=1,000$, $\langle
  k\rangle = 6$, $\gamma=3$ and $\Delta\alpha=6.0$.}
\label{map2D}
\end{figure}

We now move to validate the robustness and generality of our results against stochastic fluctuations, and/or parameter mismatches. For that purpose, we implement the electronic network schematically shown in Fig.~3. The experiment consists of six piecewise R\"ossler circuits operating in the chaotic regime, which are labeled as $N1$, $N2$, \dots, $N6$. The details of the circuit construction, as well as the qualitative equivalence of the experiment with the model of Eq. (1) are available in the literature \cite{Pisarchik06,Pisarchik09}.

The circuits are arranged in a star-like configuration, which, on its turn, represents {\it the maximally heterogeneous structure} available for small ensembles. All the chaotic oscillators have the same internal $R_{\mbox{\scriptsize exp}}$ [the experimental equivalent of the parameter $R$ in Eq. (1)], to ensure that they work in an almost equal dynamical regime. By a fine tuning of the values of the capacitors, circuits are configured such that the central node $N1$ oscillates with a mean frequency of $3,333$~Hz, and the leaves nodes $N2$, \dots, $N6$ are set with frequencies in the range of $2,240$~$\pm$~$200$~Hz. Notice that, due to the experimental variability, the frequencies of the oscillators suffer also from unavoidable dispersions.

\begin{figure}
\includegraphics[scale=0.35]{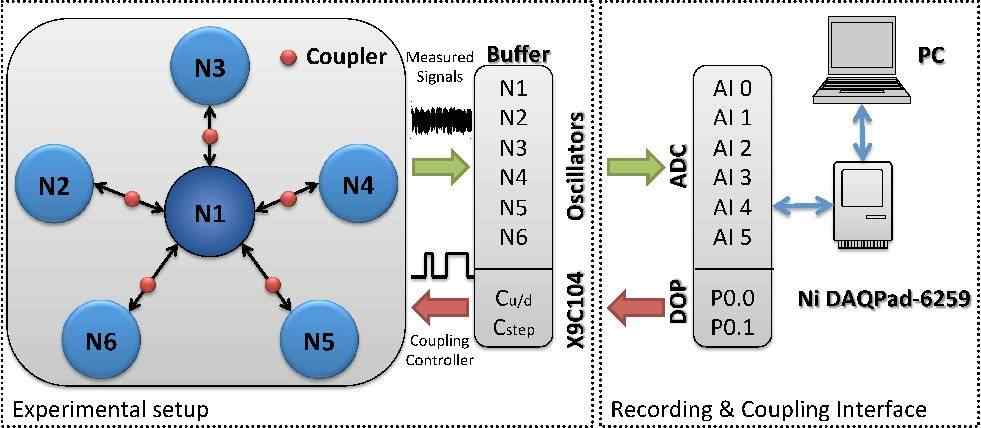}
\caption{(Color online). Sketch of the experimental setup.
Six   R\"ossler circuits (blue nodes) are connected in a star
configuration. The bidirectional coupling is adjusted by means of
five digital potentiometers X9C104 (red nodes) whose parameters
$C_{u/d}$ (value of the resistance) and $C_{\mbox{\scriptsize step}}$
(increment of the resistance at each step) are controlled
by a digital signal coming from a DAQ Card. The outputs of the six
circuits are sent to a set of voltage followers that act as a buffer and,
then, sent to the analog ports (AI~0, AI~1, \dots, AI~5)
of the same DAQ Card. The whole experiment is controlled from a PC with a Labview Software.}
\label{setup}
\end{figure}

The experiment is performed by coupling simultaneously the six oscillators (blue nodes) with the same coupling parameter $d$. We use five digital potentiometers (red nodes) in order to warrant that the parameter $d$ is changed simultaneously for all nodes. Digital potentiometers (X9C104) are adjusted by a digital signal coming from ports P0.0 and P0.1 of a NI Instruments DAQ Card (NI DAQPad-6259). The output of each circuit is connected to a voltage follower that works as a buffer. Next, signal is acquired by the analog ports (AI~0, AI~1, \dots, AI~5) of the same DAQ Card, and recorded in a PC for further analysis. The incoming signal of the analog inputs (ADC) and the signal sent through the digital outputs (DOP) are controlled and recorded by a Labview Software.

Once the data are stored, processing of them is made to obtain the equivalent instantaneous phases. These latter quantities are calculated by defining the instantaneous phase $\phi_i(t)$ of each oscillator $i$ as $\phi_i(t)=\ 2 \pi l_i + 2 \pi \frac{t-t_{l_i}}{t_{l_i+1}-t_{l_i}}$ in each interval $t_{l_i} \leq t < t_{l_i+1}$, where $t_{l_i}$ is the instant at which the $l_i^{th}$ crossing of the $i^{th}$ oscillator with its Poincar\'e section occurs. It is crucial to remark that, for phase coherent systems, such a measure of instantaneous phase is fully equivalent to the geometrical phases used in the numerical data \cite{physrepsincro}.

In Fig.~4 we report the experimental synchronization diagram for three values of the parameter $R_{\mbox{\scriptsize exp}}$, both for forward and backward variations of the coupling strength $d$. As in the numerical case, the dynamical regime tuned by $R_{\mbox{\scriptsize exp}}$ determines whether the transition is of the first or second order. The former case can be identified by the sharp increase of the synchronization parameter $S$, and the existence of a hysteresis. Interestingly, as can be seen in Fig. 2 for the two larger values of $R_{\mbox{\scriptsize exp}}$, this parameter also determines the range of $d$ at which the hysteresis appears (i.e.\ the width of the hysteresis window). Finally, it has to be noticed that the values of $R_{\mbox{\scriptsize exp}}$ at which the order of the transition changes display mismatches with those of the numerical trials, due to the imperfect equivalence between model and experimental setup.

\begin{figure}
\includegraphics[scale=0.6]{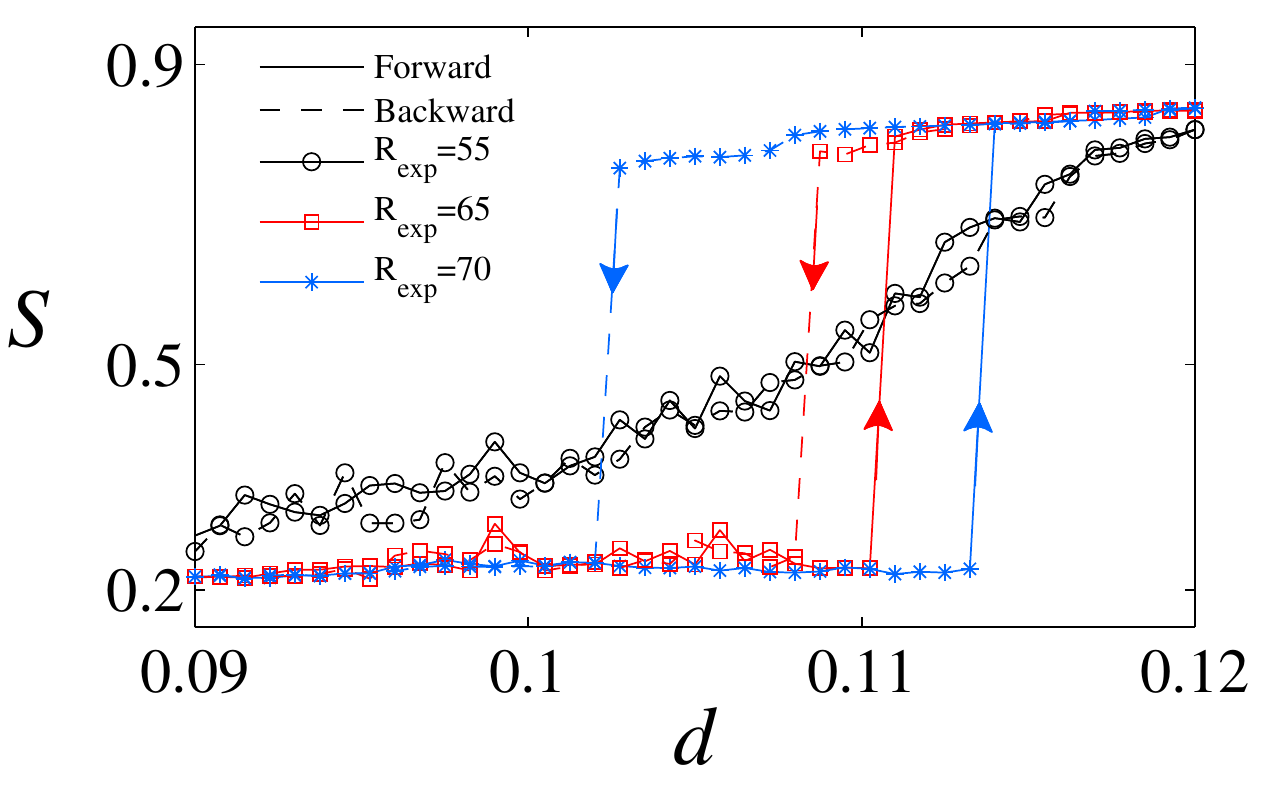}
\caption{(Color online). Phase synchronization degree
  $S$ vs.~$d$  for the star
  configuration shown in
  Fig.~3. The values of the internal resistance determining
  the chaotic state are $R_{\mbox{\scriptsize exp}}=55$ (black circles,
  second-order phase transition), $R_{\mbox{\scriptsize exp}}=65$ (red squares, first order phase transition with narrow hysteresis) and $R_{\mbox{\scriptsize exp}}=70$ (blue stars, first order phase transition with wide hysteresis).
The hysteresis in the forward (red and blue solid lines) and backward
(red and blue dashed lines) experimental trials reveals the genuine character of a first-order phase transition. The coupling strength varies at steps of $7.5 \times
  10^{-4}$ and the values of the oscillator frequencies are given in the text. }
\label{experimental}
\end{figure}

In conclusion, we have given here numerical and experimental proof of the emergence of a first order synchronization transition in a network of phase coherent chaotic oscillators. Such a discovery is the first evidence of this new paradigm of synchronization in chaotic systems. We have shown that the correlation between oscillation frequencies and structural heterogeneity is a necessary condition for such an explosive phenomenon. Furthermore, our work goes one step further by also establishing that, in chaotic systems, additional features like the specific dynamical state at which the chaotic units operate are determinant. This is of particular interest mainly for two reasons: first, real systems are noisy, and do not always work on the same regime. Therefore, whether or not a first order synchronization transition can be achieved would not depend on the specifics of the system, but on the region at which it is operating. Secondly, one does not need to fine-tune the coupling strength to get the transition, but to vary the system parameters ($R$ in our case) in a wider region. These two observations are of utmost importance when it comes to translate the uncovered mechanism into practice. We then expect that our work will open the path to the use of explosive synchronization phenomena in many relevant applications.

Work supported by Ministerio de Educaci\'on y Ciencia, Spain, through grants FIS2008-01240, FIS2009-13364-C02-01, FIS2009-07072, MTM2009-13848, and FIS2011-25167; by the Community of Madrid under project URJC-CM-2010-CET-5006; by grant 2009-SGR-838 from Generalitat de Catalunya, by Comunidad de Arag\'on (Spain) through a grant to FENOL group, and from the BBVA-Foundation within the Isaac-Peral program of Chairs. Authors acknowledge also the R\&D Program MODELICO-CM [S2009ESP-1691], and the usage of the resources, technical expertise and assistance provided by supercomputing facility CRESCO of ENEA in Portici (Italy). J.G.G. is supported by MICINN through the Ram\'on y Cajal program.

\end{document}